\documentclass[twocolumn,showpacs,preprintnumbers,aps]{revtex4}

\usepackage{graphicx}
\usepackage{dcolumn}
\usepackage{bm}

\def\bm{\boldsymbol}

\input rotate

\def\bea{\begin{eqnarray}}
\def\eea{\end{eqnarray}}
\def\ben{\begin{equation}}
\def\een{\end{equation}}

\begin{document}

\preprint{DIPC}

\title{Quantum Monte Carlo modelling of the spherically averaged
structure factor of a many-electron system}
\author{R. Gaudoin$^1$, J. M. Pitarke$^{1,2}$}
\affiliation{$^1$Donostia International Physics Center (DIPC) and
Unidad F\'\i sica Materiales CSIC-UPV/EHU,\\
Manuel de Lardizabal Pasealekua, E-20018 Donostia, Basque Country, Spain\\
$^2$Materia Kondentsatuaren Fisika Saila, Zientzi
Fakultatea, Euskal Herriko Unibertsitatea\\
644 Posta kutxatila, E-48080 Bilbo, Basque Country, Spain
}

\date{\today}

\begin{abstract}
The interaction and exchange-correlation contributions to the
ground-state energy of an arbitrary many-electron system can be
obtained from a spherical average of the wavevector-dependent diagonal
structure factor (SF). We model the continuous-k spherically averaged
SF using quantum Monte Carlo calculations in finite simulation cells.
We thus derive a method that allows to substantially reduce the
troublesome Coulomb finite-size errors that are usually present in
ground-state energy calculations. To demonstrate this, we perform
variational Monte Carlo calculations of the interaction energy of the
homogeneous electron gas. The method is, however, equally applicable
to arbitrary inhomogeneous systems.
\end{abstract}


\maketitle

Together with density-functional theory (DFT)~\cite{dft1,dft2}, Quantum Monte
Carlo (QMC) calculations belong to the bedrock of computational solid state
physics~\cite{qmc}. One major problem, the focus of this paper,
encountered in QMC solid-state applications are the Coulomb
finite-size effects. These originate in the periodic Ewald
interaction that is typically used to model the electron-electron (e-e)
Coulomb interaction in a periodic geometry~\cite{note1}. When using the Ewald
interaction, the long-range nature of the Coulomb interaction yields spurious
contributions to the interaction energy caused by the interaction of an
electron with the periodically repeated copies of its exchange-correlation
(xc) hole. Such finite-size effects are usually dealt with by increasing the
system size and monitoring the convergence of the relevant data. However, the
Coulomb finite-size errors of the interaction energy are known to scale as
$1/N$, $N$ being the number of electrons in the supercell, and convergence is
therefore slow~\cite{cep1,cep2}. An alternative involves
replacing the periodic Ewald interaction
by a "model periodic Coulomb" (MPC) interaction that converges
faster~\cite{mpc1,mpc2}.

In this Letter, we present an approach that reduces finite-size errors by
keeping the true Coulomb interaction and going to the core of the issue 
by using QMC to model the spherical average of the diagonal structure factor
(SF) of extended systems. This is in contrast to a paper~\cite{chp} we
recently learnt of that uses the (non-spherically averaged) SF to correct QMC data at long
wavelengths. Our starting point is the link between the interaction energy and
the spherically averaged wavevector-dependent diagonal SF $S_k$ of an
arbitrary many-electron system. Since large ${\bf k}$ wavevectors are related
to small electron separations, the large-$k$ behavior of QMC simulations ought
to be correct. This is not the case for the small-$k$ behavior (long
wavelengths), which in the case of a finite simulation cell 
yields spurious Ewald interactions. However, the
long-range (small-$k$) behaviour can be obtained differently, either from the
known constraint $S_{k\to 0}=0$~\cite{notenew} or from alternative
calculations based on, e.g. the random-phase approximation (RPA) which is known to
describe collective excitations correctly and becomes exact at long
wavelengths~\cite{pines}. We aim to have the best of both worlds!

The method we propose bears some resemblance to the MPC
interaction. However, while the MPC modifies the interaction, here we
keep the standard Coulomb interaction and model the QMC correlations
in a $k$ dependent way. The advantage is that our method ought to
yield improvements where others fail~\cite{mf}, e.g. the exchange hole
is known to be long-ranged, decaying as $1/r^4$. In a finite
simulation cell this results in a finite size error of the exchange
energy $\propto N^{-2/3}$. The error at large $r$, however,
corresponds to $S_{k}$ at small $k$ where it can be replaced easily by
the correct asymptotic value.

The interaction energy of an arbitrary many-electron system is usually
expressed as the sum of the Hartree energy (which in the case of an infinite
Jellium model is exactly cancelled by the Coulomb energy due to the
presence of the positive background) and the so-called xc interaction energy
$U_{xc}$. $U_{xc}$ corresponds to the attractive interaction between each electron and
its own xc hole. Starting from the spherical average $n_{xc}({\bf r},u)$ of the
xc hole density $n_{xc}({\bf r},{\bf r}')$ at ${\bf r}'$ around an electron at
${\bf r}$, one finds~\cite{lp}
\begin{equation}
U_{xc}={N\over\pi}\int dk\,(S_k-1),
\end{equation}
where $S_k$ is the spherical average of the diagonal structure factor
$S_{{\bf k},{\bf k}}$:
\begin{equation}
S_k=1+{4\pi\over N}\int d{\bf r}\,n({\bf r})\,\int du\,u^2\,
{\sin(ku)\over ku}\,n_{xc}({\bf r},u),
\end{equation}
$n({\bf r})$ being the electron density at ${\bf r}$.

\begin{figure}
\centering
\includegraphics[width=0.48\textwidth]{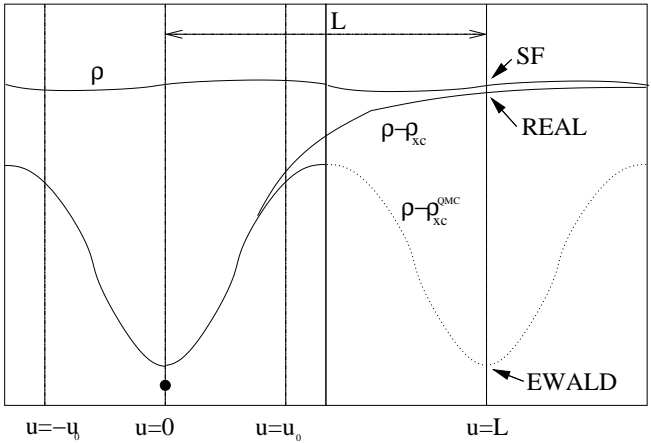}
\caption{An electron at $u=0$ inside a box of size $L$ is surrounded by a
periodic xc-hole $\rho^{QMC}_{xc}$. $U_{xc}$ is the Coulomb energy between the
electron and its xc-hole $\rho_{xc}$. Our method effectively replaces
$\rho-\rho^{QMC}_{xc}$ (EWALD in the figure) by the density $\rho$ (SF) beyond
a given cutoff $u_0$. Clearly, the interaction between an electron and its xc
hole at $u=L,2L,\ldots$ is more accurate (closer to REAL) and converges much
faster than in the standard QMC case (EWALD). Close to the boundary of the
box $\rho_{xc}$ and $\rho^{QMC}_{xc}$ diverge as both have to
integrate to $-1$. Since $\rho_{xc}$ in the SF case jumps to $0$
at $u_0$ this sum-rule is violated, but the resulting error 
can be corrected for easily (see text).}
\label {xfig}
\end{figure}

Implicitly we are considering an infinite system: $k$ is a
continuous variable. However, $S_k$ for a QMC system contains
irregularly spaced delta peaks that on integration give the QMC
$U_{xc}$. Our model (see Fig.~\ref{xfig}) assumes that correlations are
non-periodic and beyond a cutoff radius $u_0$ are due only to variations in
the density, so that beyond $u_0$ there is no contribution to $U_{xc}$. For
$u_0$, we choose the Wigner-Seitz radius $u_{WS}$ of the simulation
cell, and the structure factor $S_k$ can then be sampled directly during
the QMC run. We find:
\begin{equation}
S_k=S_k^I+S_k^{II},
\end{equation} 
\begin{equation}
\label{eq_t1}
S_k^I=\frac{1}{N}\left\langle\sum_{i\neq j}
\frac{\sin k|\bm r_j-\bm r_i|}
{k|\bm r_j-\bm r_i|}
\Theta\left(u_0-|\bm r_j-\bm r_i|\right)
\right\rangle_{QMC}, 
\end{equation}
\begin{equation}\label{eq_t2} 
S_k^{II}= -\frac{3}{2}\frac{f}{N}\sum_{\bm
q} \tilde{g}(\tilde{k},\tilde{q}) \tilde{n}_{\bm q}\tilde{n}_{-\bm q},
\end{equation}
\ben \tilde{g}(\tilde{q},\tilde{k}) =
\frac{1}{\tilde{k}\tilde{q}} \left(
\frac{\sin(\tilde{k}-\tilde{q})}{\tilde{k}-\tilde{q}} -
\frac{\sin(\tilde{k}+\tilde{q})}{\tilde{k}+\tilde{q}} \right),  
\een
using the
dimensionless quantities $\tilde{n}_{\bm q}=V n_{\bm q}$,
$\tilde{k}=ku_0$, and $\tilde{q}=|\bm q|u_0$.  $f=4\pi
u_0^3/(3V)$ is the volume
fraction of the super cell that contributes to $S_k^{I}$, 
$V$ is its total volume, and $n_{\bm q}$ denotes the Fourier transform of the
electron density $n({\bf r})$.
Apart from $n({\bf r})$ the sampling of $S_k$ only needs $|\bm r_j-\bm r_i|$ which is
readily available in most QMC codes making the SF easy to implement.
$S_k^{II}$, which is due solely to variations in
the density, 
cancels the Hartree contribution \cite{note2} to Eq. (\ref{eq_t1}).
Note that the $k$ is continuous 
even in the case of a periodic system for which the ${\bm q}$ vectors
are discrete.  This is deliberate, as the sampling
of a periodic QMC system models an extended (non periodic,
continuous $k$) system.

Equations~(\ref{eq_t1}) and (\ref{eq_t2}) do not include the entire xc hole in
the QMC sampling, as $f<1$.
As a result, our raw SF differs from zero at $k=0$.
This corresponds to the amount of the xc hole that is missed and which is
located in the corners of the simulation cell beyond the cutoff 
radius~\cite{note3}.
However, due to the periodic boundary conditions the QMC
description in these corners is unlikely to be accurate, so not much
information is lost. 
Below we show that some of the residual error
can be corrected for easily and efficiently.

We performed Hartree-Fock (HF) and Variational Monte Carlo (VMC)
calculations for the homogeneous electron gas, using the CASINO
package~\cite{casino}. The calculations employ plane-wave Slater
determinants with and without a Jastrow factor. The latter
corresponds to HF calculations where the exact result is known~\cite{pines}.
The systems we studied are non-polarised in a face-centered
cubic simulation cell with the number of electrons ranging from $2*27$ to
$2*307$ at $r_s=1$. This corresponds to a Wigner-Seitz radius $u_{WS}$ ranging
from $3.420$ to $7.689$ in atomic units, which we use throughout.
 The interaction energy is
evaluated using either our SF-based approach or the standard
Ewald interaction.  The SF is sampled at 1000 equally spaced points ranging
from $0$ to $10$.  In the case of VMC calculations, the Jastrow factor was
converged using several iterations of variance minimisation.

\begin{figure}
\centering
\includegraphics[width=0.48\textwidth]{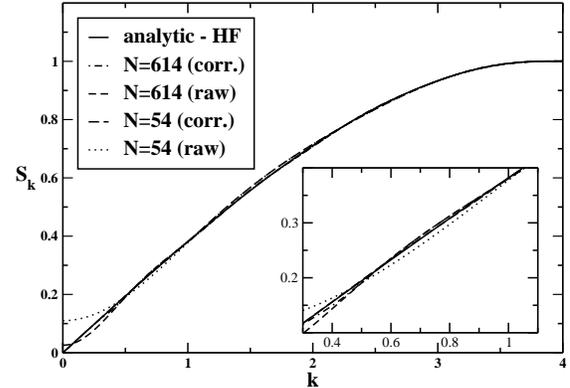}
\caption{
This figure shows the raw QMC and the
adjusted Hartree-Fock SF for the 
smallest and the largest system (2*27 and 2*307 electrons
respectively) as well as the exact result.
}
\label{hfsf}
\end{figure}

\begin{figure}
\centering
\includegraphics[width=0.48\textwidth]{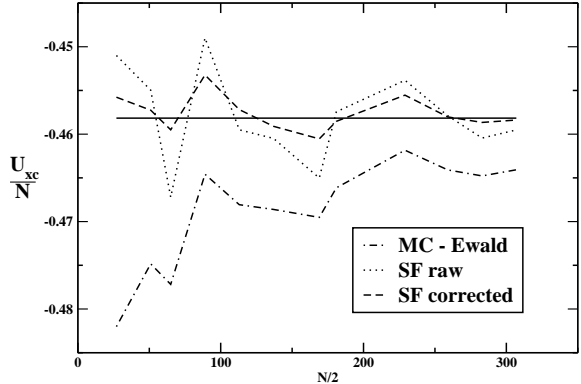}
\caption{
Convergence of the potential energy with system size (2*27 to 2*307 electrons)
using a Slater determinant only. 
The solid line is the HF result for an infinite system.}
\label{hfconv}
\end{figure}

Let us first look at a pure Slater determinant of plane waves. 
Fig.~\ref{hfsf} plots our HF calculation of the SF next to the {\it exact}
Hartree-Fock SF. Fig.~\ref{hfconv} shows the convergence of $U_{xc}$ to the
known HF (exchange) energy. The Ewald
data shows the familiar finite-size errors, 
while the SF yields an interaction (exchange)
energy that is essentially flat, consistent with the elimination of the 
Coulomb finite-size error. 

\begin{figure}
\centering
\includegraphics[width=0.48\textwidth]{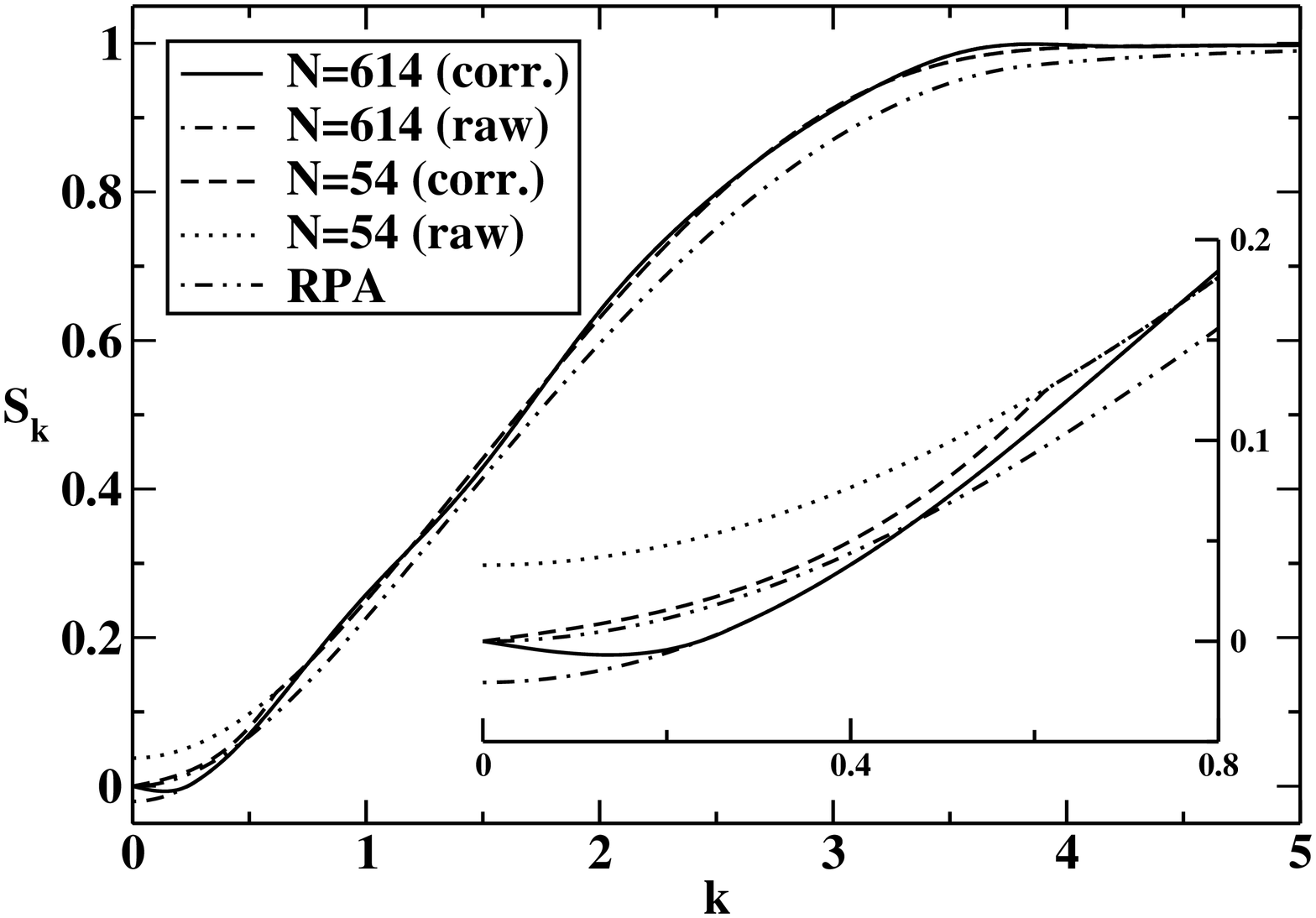}
\caption{
The figure shows the raw QMC and the adjusted Slater-Jastrow SF
for the smallest and the largest system (2*27 and 2*307 electrons
respectively).
The RPA SF is also shown. It seems incorrect for $k>0.5$. Hence, only for
systems large enough such that $0.5\gtrsim k_c$ would interpolation with the
RPA SF improve $U_{xc}$.}
\label{qmcsf}
\end{figure}

\begin{figure}
\centering
\includegraphics[width=0.48\textwidth]{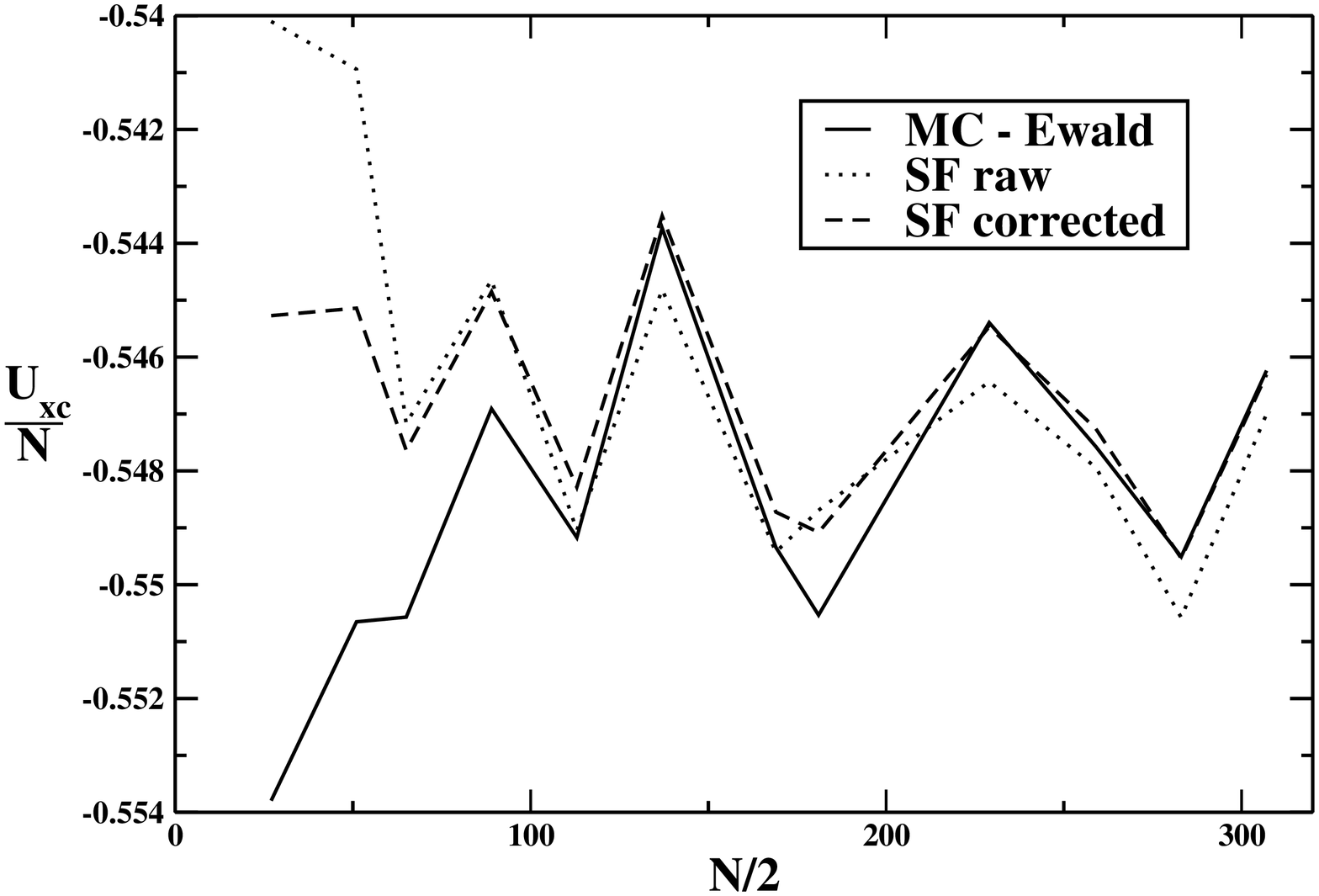}
\caption{
Convergence of the potential energy with system size (2*27 to 2*307 electrons)
using
a Slater-Jastrow wavefunction. 
The standard Ewald result is shown next to the raw and corrected SF result.}
\label{qmcconv}
\end{figure}

A similar analysis can be performed with the correlated Slater-Jastrow
many-electron wavefunction (Figs.~\ref{qmcsf} and \ref{qmcconv}). As in the
case of the HF calculation, the Ewald data exhibits finite-size errors, 
which should scale as $1/N$.
In contrast, our SF-based
calculations exhibit a systematic error with the
opposite sign. The Coulomb finite-size error has been eliminated, but we are
missing a bit of the xc hole (located at the corners of the simulation cell),
which yields a structure factor $S_k$ that differs from zero at $k=0$ 
resulting in an erroneous xc interaction energy.

We now analyse the behavior of the structure factor $S_k$ at small
$k$. As the system size
increases, one would expect the SF to improve at small
$k$. Fig.~\ref{hfsf} shows our variational HF calculation of the SF, for
two systems. The SF is essentially correct beyond a system-size
dependent minimum $k_0$. At $k<k_0$, the SF levels off and 
at $k=0$ approaches a value that equals the error in the xc
hole. The crossover $k_0$ obviously goes to zero as the system size
increases. Since our xc-hole is expected to be accurate inside a
sphere of radius $u_0$, one expects the SF to be
accurate beyond $2\pi/(2u_0)$, $2u_0$ being the characteristic length
scale of the simulation cell. Indeed, a cutoff $k_0=\pi/u_0$ seems
plausible. We have looked at the $k_0$ values at which our HF structure factor
and the exact one start to diverge markedly, and we have found $k_0\sim
4.2/u_0$. This rough estimate implies
an accurate HF xc hole within a radius $\sim 3u_0/4$ from a
given electron.
Using $k_0=4.2/u_0$ and letting our calculated HF structure factor
go to $0$ linearly at smaller $k$ produces a new {\it corrected} estimate of
$U_{xc}$ shown in Fig.~\ref{hfconv}.
There is no longer a systematic error due to the system size and
the corrected SF exhibits fluctuations that are considerably smaller than
those of the {\it uncorrected} SF. Statistical noise and a shell
structure remain, of course.

Figs.~\ref{qmcsf}-\ref{qmcconv} repeat the analysis for the correlated
Slater-Jastrow wavefunction. Here we use the VMC SF of the 
largest
system as the reference\cite{notel}. 
Interestingly, the VMC structure factor seems to remain correct
at unexpectedly small values of $k$. We have found $k_0\sim 2.1/u_0$,
corresponding to a correct xc hole up to a surprisingly large $u=1.5u_0$!
Possibly, 
$S_0=0$ poses such a strong constraint on the relatively shapeless\cite{note_shapeless}
 SF that the
SF has little choice but to be accurate at "too small" $k$ especially as
 in contrast
to the HF case, the true interacting SF  
is also quadratic for  $k\to 0$ \cite{notel}. The behavior of
the SF around $k=0$ is therefore qualitatively correct. 
 Nevertheless, a correction is needed.
We chose a simple scheme, multiplying
$S(k)$ by $k/k_0$ when $k<k_0$. We did not use the RPA for interpolation as
its region of validity seems to begin at values of $k$ smaller 
than our smallest $k_0$
(see also Fig.~\ref{qmcsf}).
 By looking at the final convergence of the xc
interaction energy $U_{xc}$ with system size, we see again that the estimates
for $U_{xc}$ are noisy but essentially flat, which is a signature of Coulomb
finite-size errors having been eliminated.

QMC being inherently statistical in nature, we close 
with a discussion of error bars. Errors of the SF at different $k$ are
correlated
and so the direct evaluation of an error for an xc interaction energy $U_{xc}$
derived in the way described here is non-trivial. 
However, observe that in a finite system the
SF-based $U_{xc}$ and the usual $U_{xc}$ coincide. Even for a
infinite system, as the simulation cell increases their values become more and
more similar as they contain similar if not
identical information. Hence, it seems reasonable to assume that the
error bars of the {\it standard} $U_{xc}$ can be used for the SF data. We have
evaluated both the SF-based and the standard $U_{xc}$ for a homogeneous
electron gas with $2*51$ electrons, for $10$
statistically independent yet identical runs averaging over $1000$ QMC steps.
The estimates for $U_{xc}$ differ by an offset, due to the different
finite-size errors, nevertheless, the estimated standard deviations are
similar: $0.00182$ using the Ewald interaction,
$0.00216$ using the uncorrected SF, and $0.00173$ using
the corrected SF. These results are consistent with an error bar of $0.0015$
derived by blocking. The estimates for $U_{xc}$ were correlated, with the
correlation between the Ewald calculation and the uncorrected and corrected
SF-based calculations being $0.64$ and $0.74$ respectively. The two SF-based
calculations had a correlation coefficient of $0.88$. 
Thus the error-bars for the standard Ewald data can be used also
 for the SF based data.

In conclusion, we have devised a new method to evaluate QMC xc interaction
energies $U_{xc}$ that do not suffer from spurious interactions of electrons
with periodic copies of their xc hole. The method is robust\cite{note_shapeless}
 and easy to implement.
Applying our method to Slater and
Slater-Jastrow type many-body wavefunctions of a homogeneous electron gas, we
have shown how to efficiently handle and eliminate 
residual Coulomb finite-size errors.
Our approach is equally applicable to arbitrary inhomogeneous
many-electron systems. Spherical averaging reduces the information
contained in any QMC system to a smooth one-dimensional curve. Each
value of $S(k)$ therefore contains more information (hence less
statistical noise) than $S(\bm k,-\bm k)$ and no spherical
self-averaging need be assumed.
 In future, we aim to apply our method to the case of Jellium surfaces
and real solids. In the case of the homogeneous electron gas that we have
considered here, we have shown that SF-based calculations of the xc
interaction energy can be improved considerably by simply letting the SF go
to the correct long-wavelength limit at $k=0$. For more complex
inhomogeneous many-electron systems it might be advantageous, however, to
splice together structure factors obtained from RPA at small $k$ and QMC at
larger $k$.

J.M.P. thanks the hospitality of the TCM Group at the Cavendish Laboratory,
where this work was initiated, and thanks R. J. Needs for enjoyable
discussions. The authors acknowledge partial support by the University of the
Basque Country, the Basque Unibertsitate eta Ikerketa Saila, the MCyT, and the
EC 6th framework Network of Excellence NANOQUANTA (Grant No.
NMP4-CT-2004-500198).


\begin{thebibliography}{[Vo]}

\bibitem{dft1} W. Kohn and L. J. Sham, Phys. Rev. {\bf 140}, A1133 (1965).
\bibitem{dft2} R. M. Dreizler and E. K. U. Gross, {\it Density-Functional
Theory. An Approach to the Quantum Many-Body Problem}, Springer, 1990.
\bibitem{qmc} W. M. C. Foulkes, L. Mitas, R. J. Needs, and G. Rajagopal, Rev.
Mod. Phys. {\bf 73}, 33, 2001.
\bibitem{note1}
 Extended systems are approximated by a finite number 
of electrons in a box with periodic boundary conditions.
\bibitem{cep1} D. M. Ceperley, Phys. Rev. B {\bf 18}, 3126 (1978).
\bibitem{cep2} D. M. Ceperley and B. J. Alder, Phys. Rev. B {\bf 36}, 2092
(1987).
\bibitem{mpc1} L. M. Fraser, W. M. C. Foulkes, G. Rajagopal, R. J. Needs, S. D.
Kenny, and A. J. Williamson, Phys. Rev. B {\bf 53}, 1814 (1996).
\bibitem{mpc2} A. J. Williamson, G. Rajagopal, R. J. Needs, L. M. Fraser, W. M.
C. Foulkes, Y. Wang, and M.-Y. Chou, Phys. Rev. B {\bf 55}, R4851 (1997).
\bibitem{chp} S. Chiesa, D. M. Ceperley, R. M. Martin, M. Holzmann,
cond-mat/0605004
\bibitem{notenew} $S_{k=0}-1=-1$ is the integrated xc hole.
\bibitem{pines} P. Nozieres and D. Pines, {\it The Theory of Quantum Liquids},
Addison-Wesley, 1989.
\bibitem{mf}  W. M. C. Foulkes, private communication.
\bibitem{lp} D. Langreth and J. P. Perdew, Phys. Rev. B {\bf 15}, 2884 (1977).
\bibitem{note2} For a homogeneous infinite many-electron system this 
term becomes the more familiar $S_k=N\delta_{k,0}$.
\bibitem{note3} Interestingly, for some Slater-Jastrow 
wavefunctions $SF(k=0)<0$ suggesting that too much electron density was
 expelled from the vicinity of an electron. This might be an artifact of the
specific Jastrow function we used.   
\bibitem{casino} R. J. Needs, M. Towler, N. Drummond, and P. Kent, {\it CASINO
version 1.7 User manual} (University of Cambridge, 2004).
\bibitem{note_shapeless}
Incidentally, due to the shapelessness of the SF 
accurate integrals can be obtained using relatively 
coarse grids.
\bibitem{notel} The correct 
structure factor of a homogeneous
electron gas goes as $S_k=k^2/2\omega_p$ for ($k\to 0$). 
 $\omega_p=(4\pi n)^{1/2}$ is the
bulk-plasmon energy and $n$ the electron
density.     

\end{thebibliography}
\end{document}